\documentclass[letterpaper, 10pt, conference]{ieeeconf}
\usepackage[autostyle]{csquotes}
\usepackage{bm}
\usepackage{amssymb}
\usepackage[linesnumbered,lined,boxed,commentsnumbered]{algorithm2e}
\usepackage{epsfig}
\usepackage{subfigure}
\usepackage{multicol, blindtext}
\usepackage{amsmath}
\usepackage{graphicx}
\usepackage{epstopdf}
\usepackage{float}
\usepackage{amsfonts}
\usepackage{color}
\usepackage{dblfloatfix}
\usepackage{multirow}%
\usepackage{fourier} 
\usepackage{array}
\DeclareMathOperator*{\argmin}{argmin}   
\DeclareMathOperator*{\argmax}{argmax}   
\usepackage{algpseudocode}
\usepackage{makecell}

\providecommand{\U}[1]{\protect\rule{.1in}{.1in}}

\newtheorem{definition}{\rm\textbf{Definition}}
\newtheorem{property}{\rm\textbf{Property}}

\IEEEoverridecommandlockouts
\title{Merging control in mixed traffic with safety guarantees: a safe sequencing policy with optimal motion control}
\date{August 2022}

\author{Ehsan Sabouni$^1$, H.M. Sabbir Ahmad$^1$, Christos G. Cassandras$^1$ and Wenchao Li$^1$
\thanks{$^1$Division of Systems Engineering and Department of Electrical \& Computer Engineering, Boston University, Boston, MA, USA, \texttt{{\small \{esabouni, sabbir92, cgc, wenchao\}@bu.edu}}}
\thanks{This work was supported in part by ONR grant N00014-19-1-2496, NSF under grants ECCS-1931600, DMS-1664644, CNS-1645681, CNS-2149511, by AFOSR under grant FA9550-19-1-0158, by ARPA-E under grant DE-AR0001282, and by NPRP grant (12S-0228-190177) from the Qatar National Research Fund, a member of the Qatar Foundation (the statements made herein are solely the responsibility of the authors).}}

\begin{document}
\maketitle
\begin{abstract}
We address the problem of merging traffic from two roadways consisting of both Connected Autonomous Vehicles (CAVs) and Human Driven Vehicles (HDVs). Guaranteeing safe merging in such mixed traffic settings is challenging due to the unpredictability of possibly uncooperative HDVs. We develop a hierarchical controller where at each discrete time step first a coordinator determines the best possible Safe Sequence (SS) which can be realized without any knowledge of human driving behavior. Then, a lower-level decentralized motion controller for each CAV jointly minimizes travel time and energy over a prediction horizon, subject to hard safety constraints dependent on the given safe sequence. This is accomplished using a Model Predictive Controller (MPC) subject to constraints based on Control Barrier Functions (CBFs) which render it computationally efficient. Extensive simulation results are included showing that this hierarchical controller outperforms the commonly adopted Shortest Distance First (SDF) passing sequence over the full range of CAV penetration rates, while also providing safe merging guarantees.

\end{abstract}

\thispagestyle{empty} \pagestyle{empty}


\section{INTRODUCTION} 

The emergence of Connected and Automated Vehicles (CAVs) has the potential to drastically impact transportation systems in terms of increased safety, as well as reducing congestion, energy consumption, and air and noise pollution \cite{li2013survey}. In particular, CAVs enable intelligent traffic management at 
conflict areas, such as intersections, roundabouts, and merging roadways, through cooperation using connectivity, all of which critically affect the performance of a traffic network \cite{MALIKOPOULOS_survey2017}. Such intelligent traffic management is usually based on formulating and solving optimal control problems for CAVs seeking to jointly minimize travel times and fuel consumption while always satisfying constraints associated to safety.

To date, most of the research involving the optimal control of CAVs has been designed assuming a $100\%$ CAV penetration rate, which is unlikely to be the case for the foreseeable future \cite{ALESSANDRINI2015145}. This creates the need for controlling CAVs in a \emph{mixed traffic} environment where CAVs co-exist with Human Driven Vehicles (HDVs). 
Since human driving is unpredictable \cite{ZHENG2020Impact} and HDV behavior can be stochastic and selfish \cite{delle2021pricing}, this can lead to congestion and safety issues \cite{mahdinia2021integration},\cite{bhavsar2017risk},\cite{al2021impacts}. 

Efforts to model HDV behavior in mixed traffic commonly assume that HDVs maintain constant speed in free-flow traffic and adopt one of several car-following models, e.g., \cite{olstam2004comparison}, \cite{munigety2016towards}. In \cite{Arvin2020Mixedtraffic}, \cite{yao2020decentralized}, the safety of CAVs in intersections with mixed traffic is evaluated with the HDV behavior modeled by the Wiedemann car-following model. Lane change maneuvers in a mixed traffic setting were also studied in \cite{li2022simulation}. However, these assumptions do not capture the full range of possible HDV behaviors, which can be highly variable as drivers may react differently to different situations.
Ultimately, the goal is to develop control and coordination algorithms for CAVs that can handle the full range of HDV behaviors in mixed traffic conditions, while ensuring safe and efficient traffic flow. Towards this goal, a bi-level optimization problem is formulated in \cite{MixedtrafficSurverymergingZhao2019} to determine a merging sequence and control input for CAVs in merging roadways with mixed traffic so as to achieve safe merging, assuming an intelligent car-following model for HDVs. In \cite{LiaoGameTheory2022} a game theoretic approach is proposed to improve the performance metrics of the network, but safety guarantees are not yet provided. In \cite{mixedmergingLiu}, the safety of merging in a mixed traffic scenario is considered, but sequencing schemes are not considered. 

Focusing on merging roadways, the problem of developing schemes for both control and coordination of CAVs in mixed traffic, while also providing safety guarantees, remains a challenging one.
In prior work \cite{sabouni2022optimal}, the merging of a single AV and multiple HDVs was studied and it was shown that merging the AV after all HDVs or ahead of all HDVs can be optimal under certain conditions. Motivated by this idea, we consider arbitrary penetration rates of CAVs and seek solutions to the optimal merging problem that can guarantee safety under arbitrary HDV behavior.

The main contribution of this paper is to introduce the concept of \emph{Safe Sequencing} (SS) of merging CAVs as a technique for suppressing (and possibly eliminating) the number of cases where a CAV merges in front of an HDV; it enforces instead a sequence such that each CAV merges in front of another CAV, hence the two can safely cooperate. This suppresses the risk of safety violations due to the unpredictability of merging HDVs when they are faced by CAVs that conflict with them. To make effective use of this concept, we decompose the complex nonlinear mixed integer programming problem which defines the joint coordination and motion control of 
CAVs in mixed traffic at merging roadways. First, an upper-level controller at a local coordinator determines the best possible \emph{safe} sequence at each discrete time step. Next, this information is provided to each CAV which solves a lower-level decentralized optimization problem to jointly minimize travel time and energy subject to hard safety constraints dependent on the given safe sequence. 
We use a Model Predictive Controller (MPC) subject to constraints based on Control Barrier Functions (CBFs), thus exploiting their forward invariance property and the computational efficiency resulting from the linearity of the CBF-based constraints.
We empirically show that the proposed scheme outperforms the commonly adopted Shortest Distance First (SDF) passing sequence while also providing safety guarantees that SDF sequencing lacks. Moreover, this is achieved \emph{without any knowledge of human driving behavior}
and despite the fact that SS is expected to be conservative by prioritizing safety over optimality in sequencing CAVs and HDVs. The main reason is that by grouping CAVs through SS we reduce the number of instances where vehicles need to accelerate or decelerate hard as they approach the merging point and, on average, this overtakes the conservativeness of SS. In addition, the SS approach, in conjunction with the MPC/CBF control of CAVs, emphasizes the benefit of CAV presence even at low penetration rates; thi is unlike SDF sequencing schemes that require a dominant CAV flow.

The paper is organized as follows. In Section II, we define the merging problem framework and formulate a challenging problem for sequencing and motion control of all CAVs along with discussing the issues to set the stage for the subsequent sections. Section III describes hierarchical control for sequencing and control of CAVs in the sequencing zone. Firstly we present an algorithm to determine a safety prioritized merging sequence centrally, followed by our decentralized MPC-CBF controller for CAVs used to follow that sequence in the sequencing zone. Section IV, discusses the low level controller for the CAV in the awareness zone. Finally, in Section V, simulation results are presented that validate our presented control and coordination framework for varying CAV penetration rates along with specific scenarios created to highlight the merit of our presented framework.

\section{Problem Formulation}

\begin{figure}
\centering
\includegraphics[scale=0.5]{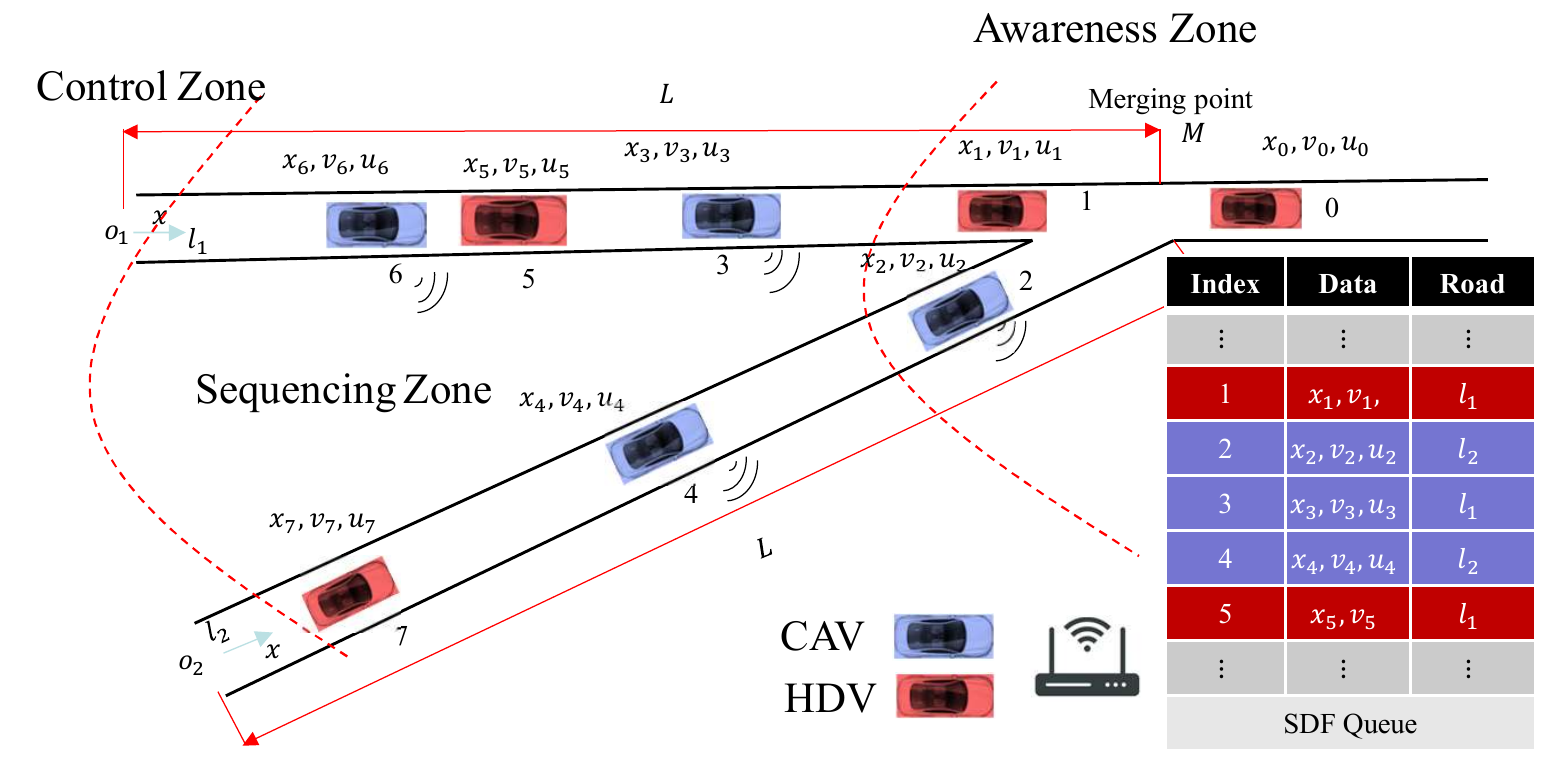} 
\caption{The merging problem }
\label{fig:merging}
\end{figure}

\label{sec:problem}

We consider the merging problem where traffic arrives from two roads and merges at a single Merging Point (MP) $M$ (see Fig. \ref{fig:merging}) where a collision might occur. The roads are indexed by $r \in \{1,2\}$, with 1 usually reserved for the ``main road''.
In this paper, we consider traffic to consist of both HDVs and CAVs (red and blue vehicles, respectively, in Fig. \ref{fig:merging}). A \textit{Control Zone} (CZ) is defined to be an area within which vehicles can communicate their states, or their states can be estimated using available sensing equipment in the road network by a roadside (RSU) coordinator unit. We assume that only CAVs are equipped with connectivity, hence, can communicate with the coordinator; HDVs have no connectivity capability. 

Let $F(t)$ be the set of the unique indices of all vehicles inside the CZ at time $t$. We will make use of two partitions of $F(t)$: $(i)$ Let $F_C(t)$ contain the indices associated with CAVs in the CZ, and $F_H(t)$ those associated with HDVs inside the CZ at time $t$ (i.e., $F_C(t) \cup F_H(t) = F(t), \ \ F_C(t) \cap F_H(t) = \emptyset  $). 
$(ii)$ Let $F_1(t)$ and $F_2(t)$ be the sets of indices of all vehicles inside the CZ at time $t$ traveling on roads 1,2 respectively (i.e. $F_1(t) \cup F_2(t) = F(t), \ \ F_1(t)\cap F_2(t) = \emptyset$).
Let $N(t)$ be the
cardinality of $F(t)$ and similarly we define $N_H(t)$, $N_C(t)$, $N_1(t)$ and $N_2(t)$ as the cardinalities of $ F_H(t)$, $F_C(t)$, $F_1(t)$ and $F_2(t)$ respectively. 
Thus, if a vehicle arrives at time $t$, it is assigned the
index $N(t)+1$ (i.e., either $N_H(t)+1$ or $N_C(t)+1$). The index $0$ is reserved for a vehicle that has just left the CZ (see Fig.\ref{fig:merging}). All vehicle indices in $F(t)$ decrease by one when a vehicle passes over the MP and the vehicle whose index is $-1$ is dropped. 

The coordinator maintains a table as shown in Fig. \ref{fig:merging} containing the states of all vehicles $i \in F(t)$.
This state information at time $t$ can be obtained by using cameras installed in the vicinity of the MP if $i \in F_H(t)$ or through direct connectivity when $i \in F_C(t)$.

The vehicle dynamics are assumed to be of the form
\begin{equation} \label{VehicleDynamics}
\left[
\begin{array}
[c]{c}%
\dot{x}_{i}(t)\\
\dot{v}_{i}(t)
\end{array}
\right]  =\left[
\begin{array}
[c]{c}%
v_{i}(t) \\
u_{i}(t) 
\end{array}
\right],  
\end{equation}
where $x_{i}(t)$ denotes the distance from the origin at which car $i$ arrives, $v_{i}(t)$ denotes the velocity, and $u_{i}(t)$ denotes the control input (acceleration). 
\\
\textbf{Objective 1} (Minimize travel time): Let $t^0_i$ and $t^m_i$ denote the time that vehicle $i \in F(t)$  enters the CZ from either road and exits through the MP.
We wish to minimize the total travel time $t^m_i-t_i^0$ of each vehicle.
\\
\textbf{Objective 2} (Minimize energy consumption): We wish to minimize the total energy consumption of each vehicle in the CZ
where the energy consumption model $E_i$ is expressed as:
\begin{equation} \label{energy}
E_{i}\big(t_{i}^{m},u_{i}(t)\big)=\int_{t_{i}^{0}}^{t_{i}^{m}}\mathcal{C}(u_{i}(t))dt,
\end{equation}
where $\mathcal{C}(\cdot)$ is a strictly increasing function of its argument. For simplicity, we limit ourselves to the case where $\mathcal{C}(u_i(t))=\frac{1}{2}u_i^2(t)$ (more accurate energy models can also be considered as in \cite{XIAO2021109592})

While in the CZ, vehicles are subject to the following constraints:\\
{\bf Constraint 1} (Vehicle limitations):  There are constraints on the speed and acceleration for each $i\in F(t)$:
\begin{equation}
\label{VehicleConstraints1}%
\begin{aligned} v_{min} \leq v_i(t)\leq v_{max}, ~~~\forall t\in[t_i^0,t_i^m]\end{aligned} 
\end{equation}
\begin{equation}
\label{VehicleConstraints2}%
\begin{aligned} u_{{min}}\leq u_i(t)\leq u_{{max}}, ~~~\forall t\in[t_i^0,t_i^m] \end{aligned} 
\end{equation}
where $v_{\max}> 0$ and $v_{\min} \geq 0$ denote the maximum and minimum speed allowed
in the CZ, $u_{{\min}}<0$ and $u_{\max}>0$ denote the minimum and maximum
control, respectively.
\textbf{Constraint 2} (Rear-end safety constraint): Let $i_p \in F(t)$ denote the index of the vehicle which physically immediately precedes CAV $i \in F_C(t)$ (if one is present). We require that the distance $z_{i,i_p}(t):=x_{i_p}(t)-x_{i}(t)$ be constrained by:
\begin{equation}
\label{rear-end} 
z_{i,i_p}(t) - \varphi v_{i}(t) - \delta \geq 0, \text{ \ }\forall t\in\lbrack
t_{i}^{0},t_{i}^{m}], 
\end{equation}
where $\varphi$ denotes the reaction time and $\delta$ is a given minimum safe distance. If we define $z_{i,i_{p}}$ to be the distance from the center of $i$ to the center of $i_{p}$, then $\delta$ depends on the length of these two vehicles\\
{\bf Constraint 3} (Safe merging): Whenever a CAV crosses the MP, a lateral collision is possible and there must be adequate safe space for the CAV to avoid it, i.e.,
\begin{equation}
\label{SafeMerging_tim}
z_{i,i^+}(t_{i}^{m})-\varphi v_{i}(t_{i}^{m})-\delta\geq 0,
\end{equation}
where $i^+ \in F(t)$ is the index of the vehicle (CAV or HDV) that may collide with CAV $i$ at $M$ as shown in Fig. \ref{fig:merging}. The determination of $i^+$ depends on the driving behavior of HDVs and the policy adopted for sequencing CAVs through the CZ.
Note that (\ref{SafeMerging_tim}) is defined for $t = t_{i}^{m}$ only. In order to make use of $z_{i,i^+}(t_{i}^{m})-\varphi v_{i}(t_{i}^{m})-\delta$ as a CBF, we need to transform it to a continuously differentiable form in $t$ using the method adopted in \cite{xiao2023safe}:
\begin{align} \label{SafeMerging} 
x_{i^+}(t) - x_{i}(t) - \Phi\big(x_i(t)\big) v_{i}(t)-\delta \geq 0,
\end{align}
where, $\Phi\big(x_i(t)\big)$ is any continuously differentiable function in $x_i$ such that $\Phi\big(x_i(t^0_i)\big) = 0$ and $\Phi\big(x_i(t_i^m)\big) = \varphi$. The linear function $\Phi(x_i(t))=\varphi x_i(t)/L$, which satisfies both conditions, is commonly used, 
where $L$ is the length of road traveled by the CAV from its entry to the CZ to the MP. Observe that constraints 1-3 defined above suffice for safe merging when all vehicles are CAVs. 
Mixed traffic, however, requires an additional safe merging constraint due to possibly uncooperative HDVs:\\
\textbf{Constraint 4}: Similar to the definition of $i^+$, let $i^-$ be the vehicle that CAV $i$ has to merge ahead of. Therefore, this CAV has to have enough space to avoid any collision:
\begin{equation} 
\label{merging_2_tim} 
z_{i^-,i}(t^m_i) - \varphi v_{i^-}(t^m_i) - \delta \geq 0,
\end{equation}
For CAV $i$, this new constraint ensures that the trailing HDV (if present) has enough space to avoid any collision; otherwise, CAV $i$ cannot merge ahead of it and must yield. Note that if $i^- \in F_C(t)$ this constraint is rendered redundant as CAV $i$ is already a merging constraint for CAV $i^-$. Similar to (\ref{SafeMerging}), we transform this constraint to obtain \begin{equation}
\label{SafeMerging_2} 
x_{i}(t) - x_{i^-}(t) - \Phi\big(x_{i^-}(t)\big) v_{i^-}(t)-\delta \geq 0,
\end{equation}

Before proceeding to formulate the joint sequencing and motion control problem, we introduce some useful notation.
Given $r \in \{1,2\}$ as the index of a specific road, let $\Bar{r}$ denote the other road. Regarding the sequence $F(t)=\{1,\ldots,j,\ldots,N(t)\}$, the prefix and suffix sub-sequences of $j$ are pertinent. Accordingly, we define
$F^-_j(t) = \{1,2,\ldots,j\}$ and $F^+_j(t) = \{j+1,\ldots,N(t)\}$.


\textbf{Optimal sequencing and motion control problem.} Our goal is to determine a control law for all CAVs $i \in F_C(t)$ in a centralized fashion so as to achieve objectives 1-2 subject to constraints 1-4 for each $i \in F_C(t)$ and vehicle dynamics (\ref{VehicleDynamics}) with the indices $i^+$ and $i^-$ in (\ref{SafeMerging_tim}) and (\ref{merging_2_tim}) respectively determined by a sequencing policy.
Without loss of generality, the optimization problem is formulated for CAV $i \in F_r$ (the formulation is the same for any CAV $i \in F_{\bar{r}}$). 
In this problem, we use the weight $\alpha\in[0,1]$ to construct a convex combination of time and energy metrics.
Defining the controllable vectors 
$\textbf{u}_c = \big\{u_i| ~i \in F_r(t) \cap F_C(t)\big\} $, 
$\boldsymbol{t}_c^m = \big\{t_i^m| ~i \in F_r(t)\cap F_C(t)\big\}$ and 
$\boldsymbol{y}_j = \big\{y_{i,j}| ~i \in F_r(t)\cap F_C(t), ~j \in F_{\bar{r}} \big\}$,
this nonlinear mixed integer optimization problem is formulated as
\begin{align} \label{eqn:tot_car_obj}
 &\min_{\substack{\boldsymbol{u}_c,\boldsymbol{t}_c^m,\boldsymbol{y}_j}} J(\boldsymbol{u}_c,\boldsymbol{t}_c^m)= \sum_{i=1}^{N(t)}\bigg(\alpha(t_i^m-t_i^0)+(1-\alpha)E_i\bigg) \\  
\textnormal{s.t.:} \ \  
  & v_{min} \leq v_i(t)\leq v_{max}, ~~u_{max} \leq u_i(t)\leq u_{max} \\  
  & z_{i,i_p}(t)\geq\varphi v_{i}(t)+\delta,  \\ 
& z_{i,j}(t_{i}^{m})+B(1-y_{i,j})\geq\varphi v_{i}(t_{i}^{m})+\delta,  ~\forall j \in F_{\Bar{r}}(t)  \\
& z_{j,i}(t_{i}^{m})+By_{i,j}\geq\varphi v_{j}(t_{i}^{m})+\delta,  ~\forall j \in F_{\Bar{r}}(t), \\
& \sum_{j \in F_{\Bar{r}}} y_{i,j} = |F^+_j(t)|\\ 
& y_{i,j} \in \{0,1\}
\end{align}
where the binary decision variables $y_{i,j}$ correspond to vehicles with index $j$ in a different road than CAV $i$. 
Constraints (11) and (12) are safe merging constraints for CAV $i$ which cannot be both satisfied at the same time. Once a specific $j$ for CAV $i$ to merge with is selected, the vehicles in the other road will be divided into two groups: $F^-_j(t)$ such that CAV $i$ merges ahead of them, and $F^+_j(t)$ behind which CAV $i$ merges. For the first group, constraint (11) is not required, while for the second group, constraint (12) is not required. Thus, the constant $B$ is chosen as a sufficiently large positive real number so that only one constraint can be satisfied at a time, while the other is automatically fulfilled. Constraint (13) is needed to ensure the uniqueness of vehicle $j$ for each CAV $i$. 

 There are two main challenges in this problem: 1) It requires a central computing resource to solve it and send the control input to the CAVs. Due to the complexity of the mixed integer program, solving it is computationally prohibitive and infeasible for real-time applications. 
 2) This problem is complicated by the fact that obtaining an expression for each HDV's travel time $t_i^m$ and energy consumption $E_i$ for $i \in F_H(t)$ as a function of the CAV's states and control input is a difficult task. 
 
 It is, therefore, natural to decompose this problem into its sequencing control and individual CAV motion control components.
 The former is solved by the coordinator to determine $j$ for each CAV $i \in F_C(t)$, hence the merging sequence. The latter is then solved in a decentralized fashion by each CAV $i$ to derive the control input $u_i(t)$ for each $i \in F_C(t)$ so as to minimize objectives 1-2. 
 In the sequel, we first present an upper-level controller which determines a \emph{Safe Sequencing} (SS) scheme by prioritizing sequences that guarantee safe merging. This controller only applies to vehicles located in a subset of the CZ termed \emph{Sequencing Zone} (SZ) denoted by $F^{SZ}(t)$. This is followed by an \emph{Awareness Zone} (AZ), an area close to the MP with length $L_{AZ}$ (see Fig. \ref{fig:merging}) containing vehicles $i \in F^{AZ}(t)$. Within the AZ, vehicles can directly observe and interact with each other and there is no longer adequate time for controlled sequencing; consequently, only the lower-level controller is considered for CAVs within the AZ. 

 While CAVs are still within the SZ, the lower level controller receives the safe sequencing information from the coordinator and employs Model Predictive Control (MPC) onboard each CAV with constraints based on Control Barrier Functions (CBFs) used to guarantee all safety constraints; this exploits the CBF forward invariance property \cite{xiao2023safe}. We assume that time is discretized and at each discrete time step $t$ where  $ k = 1,2,\ldots$ the sequencing controller is invoked, followed by the optimal motion controllers for each CAV operating within the given SS scheme.

\section{Hierarchical control in the Sequencing Zone}
Although First In First Out (FIFO) is a popular sequencing policy that performs well on symmetric merging configurations with $100\%$ CAV traffic, such \emph{fixed} policies fail in mixed traffic since HDVs do not conform to any such policy and tend to behave selfishly. We use the Shortest Distance First (SDF) policy to find candidate $i^+$ and $i^-$ introduced in constraints 3 and 4. We call them candidates since the unpredictability of HDV behavior cannot ensure safe merging under this policy. The purpose of the upper-level controller is to define instead a \emph{safe} policy derived from SDF.

\subsubsection{\textbf{Upper level control: safe sequencing}}
At every time step $t$, the coordinator determines a \emph{safe} (precisely defined below) merging sequence in a centralized manner. This is accomplished in the following steps.

\textbf{Step 1: Generate the SDF sequence.}
We define the SDF sequence at $t$, denoted by
$\boldsymbol{s}^0(t)$ ($t$ will be henceforth omitted for simplicity), as the sequence of vehicles such that all $i,j \in F^{SZ}(t)$ satisfy:
\begin{equation} \label{SDF}
L-x_i(t) <  L-x_j(t)  \Leftrightarrow  \boldsymbol{s}^0(i) <\boldsymbol{s}^0(j),
\end{equation}
where $L-x_i(t)$ is the remaining distance of $i$ from the MP and 
$\boldsymbol{s}^0(i)$ denotes the position of CAV $i$ in the sequence. 

For any $i \in \boldsymbol{s}^0$, let us identify the vehicles that $i$ must merge ahead and behind of. These are similar to ${i}^+$ and ${i}^-$ in (\ref{SafeMerging_tim}), (\ref{merging_2_tim}) and, for the SDF sequence, are denoted by $\hat{i}^+$ and $\hat{i}^-$.
Letting $\textnormal{ind}[\boldsymbol{s}^0(i)]$ denote the index of the vehicle at position $\boldsymbol{s}^0(i)$ in $\boldsymbol{s}^0$:
\begin{equation} \label{i-_search}
\hat{i}^- = \min_{j>0} \biggl\{\textnormal{ind}\big[\boldsymbol{s}^0(i)+j\big]\biggl\} ~ \textnormal{s.t.} ~  i \in F_r \implies \textnormal{ind}\big[\boldsymbol{s}^0(i)+j\big] \in F_{\bar{r}}
\end{equation}
\begin{equation} \label{i+_search}
\hat{i}^+ = \min_{j>0} \biggl\{\textnormal{ind}[\boldsymbol{s}^0(i)-j]\biggl\} ~ \textnormal{s.t.} ~  i \in F_r \implies \textnormal{ind}\big[\boldsymbol{s}^0(i)-j\big] \in F_{\bar{r}}
\end{equation}
As an example, if $\boldsymbol{s}^0 = \{1,2,3,4,5,6,7 \}$
in Fig. \ref{fig:merging}, then $\hat{4}^- = 5$.

\textbf{Step 2: Generate safe sequences.}
In the presence of HDVs, the SDF sequencing policy can lead to safety issues. For example, when ${i}^-$ for CAV $i$ is a
HDV (e.g., $4^- = 5$ in Fig. \ref{fig:merging}, safe merging cannot be ensured for two main reasons: $(i)$ there is no assumption on the HDV behavior which may not be cooperative, and 
$(ii)$ even if cooperation is assumed, there is no way to ensure the vehicles follow the SDF policy since they are on different roads and may not detect each other (this is only true in the AZ, which is why it is treated differently).
Therefore, one way to prevent this problem is for CAV $i$ to \emph{always merge ahead of another CAV}, i.e., set ${i}^- \in F_C(t)$. However, this may be overly conservative since there are cases when ${i}^- \in F_H(t)$ but the relative distance from the MP between the HDV and CAV is positive and large, 
In this case, yielding to make room for the HDV at the MP is excessively conservative. This can be avoided by incorporating a distance threshold in the determination of $i^-$ and $i^+$. Thus, we extend the feasible set for $i^+, i^-$ to $\bigl\{\textnormal{CAV}, \textnormal{HDV}, \emptyset \bigl\}, \ \forall i \in F^{SZ}_C(t)$ where $\{\emptyset\}$ signifies that there is no vehicle within the preset threshold; in other words, if $i^- = \emptyset$, then CAV $i$ is unconstrained in planning an optimal merging trajectory and needs to only consider $i^+$. This is formalized as follows. \\
\textbf{Merging pair determination}: 
Recalling the safe merging constraints (\ref{SafeMerging}), (\ref{SafeMerging_2}), we
define $\Delta^-_i(j) = x_{i}(t) - x_{j}(t) - \Phi\big(x_{j}(t)\big) v_{j}(t)-\delta $ and  $\Delta^+_i(j) = x_{j}(t) - x_{i}(t) - \Phi\big(x_i(t)\big) v_{i}(t)-\delta$.
As long as $\Delta^-_i(j) < 0$, we use 
$\hat{i}^-$ from (\label{i-_search}) as $i^-$; otherwise we set $i^- = \emptyset$. The same applies to $i^+$ and we have:
\begin{align}\label{i-i+}
    i^- = \begin{cases}
     \hat{i}^- & \text{if $\Delta^-_i(\hat{i}^-) < 0$}, \\
    \emptyset & \text{otherwise},
    \end{cases}, \ \ \
    i^+ = \begin{cases}
     \hat{i}^+ & \text{if $\Delta^+_i(\hat{i}^+) < 0$}, \\
    \emptyset & \text{otherwise},
    \end{cases}
\end{align}

\begin{definition} \label{safe_sequence}
(Safe sequence) A merging sequence of $i \in F^{SZ}(t)$ is safe if $i^-\in \{CAV, \emptyset \}$ for all $i \in F^{SZ}_{C}(t)$.%
\end{definition}

This definition is irrespective of $i^+$  since CAVs are cooperative and can invariably make room for other HDVs or CAVs. Thus, in a safe sequence, every CAV that merges ahead of another CAV their mutual cooperation ensures safe merging; if $i^- = \emptyset$, then any vehicle following $i$ is located further than the preset threshold and \eqref{i-i+} can be satisfied to ensure safe merging. 

Using Definition \ref{safe_sequence}, at the conclusion of \textbf{Step 2} a set safe sequences is generated as $\mathcal{S}_S$. 

\textbf{Step 3: Generate minimally disruptive safe sequences.}
It is likely that the set $\mathcal{S}_S$ contains multiple safe sequences. In order to differentiate among them, we introduce the concept of \emph{disruption}, defined as follows:
\begin{equation} 
    D(\boldsymbol{s},\boldsymbol{s}^0) = \sum_{i = 1}^{N(t)} \mathbb{1} \big[\boldsymbol{s}(i) \neq \boldsymbol{s}^0(i)\big] 
\end{equation}
 where $D(\boldsymbol{s},\boldsymbol{s}^0)$ measures the disruption caused by a safe sequence $\boldsymbol{s}$ relative to the SDF sequence $\boldsymbol{s}^0$ as one with the least number of changes in indices compared to $\boldsymbol{s}^0$ ($\mathbb{1}(\cdot)$ is the indicator function). 
Thus, the subset of safe sequences with minimal disruption is determined:
 \begin{equation}
    \centering
    \mathcal{S}_O = \argmin_{\boldsymbol{s} \in \mathcal{S}_S} D(\boldsymbol{s},\boldsymbol{s}^0). \nonumber
\end{equation}

 
\textbf{Step 4: Determine the optimal safe sequence $\boldsymbol{s}_F$.}
If $|\mathcal{S}_O| = 1$, the optimal safe sequence is obtained in \textbf{Step 3}. If $|\mathcal{S}_O| > 1$, then we choose the sequence $\boldsymbol{s}_F$ that gives higher priority to vehicles located on road $r \in \{1,2\}$ with higher average velocity. Thus, if $r_1$ is the road with a higher average velocity, the optimal safe sequence can be formulated as follows:  
\begin{equation}
    \boldsymbol{s}_F = \argmin\limits_{\boldsymbol{s} \in \mathcal{S}_O}\Bigg({\mathlarger{\sum}\limits_{i \in F_{r_1}}i} - {\mathlarger{\sum}\limits_{j \notin F_{r_1}}j}\Bigg),
\end{equation}
It is straightforward to establish the following two properties of safe sequences.
\begin{property}
    The optimal safe sequence $\boldsymbol{s}_F$ satisfies all original rear-end safety constraints in \eqref{rear-end}.
\end{property}

\begin{property}
    There always exists a safe sequence $\boldsymbol{s}$.
\end{property}

\begin{equation}
    {\boldsymbol{s}_S = \{\boldsymbol{s}^{0}i^+,i^-F^{SZ}_{\Bar{r}},iF^{SZ}_r\}}
\end{equation}

\begin{algorithm}[h]
\SetKwInOut{Input}{Input}\SetKwInOut{Output}{Output}
\SetKwComment{comment}{\#}{}
\Input{$\boldsymbol{x_{i}}(t) \ \forall i \in F^{SZ}(t)$} 
\Output{The new set of indices $\mathcal{S}_F$ is safe for the CAVs.}

$v^{a}_{1}$ = average velocity of cars in main road $(r = 1)$ \\
$v^{a}_{2}$ = average velocity of cars in side road $(r = 2)$\\

Compute the index set based on $\boldsymbol{s}^0$ where the tuple $(i^-, i^+), \forall i \in F^{SZ}_{C}(t)$ is computed using \eqref{i-i+} \\

\If{$\exists i \in F^{SZ}_C(t) \cap \boldsymbol{s}^0 \ s.t. \ i^- \in F_H(t)$}
    {
        Generate all admissible safe sequences at time $t$ denoted as $\mathcal{S}_S$ \\
        {
        Determine the sequences that causes least disruption compared to $\mathcal{S}_S$ as follows: \\
        }
        \begin{equation}
            \centering
            \mathcal{S}_{O} = \argmin_{\boldsymbol{s} \in \mathcal{S}_S} D(\boldsymbol{s},\boldsymbol{s}^0) \nonumber
        \end{equation}
        
        \If{$v^{a}_{1} \geq v^{a}_{2}$}
        {
            Choose the sequence from $\mathcal{S}_O$ using the following: $\boldsymbol{s}_F= \argmin\limits_{\boldsymbol{s} \in \mathcal{S}_O}{\mathlarger{\sum}\limits_{i \in F_1}i} - {\mathlarger{\sum}\limits_{j \in F_2}j}$ 
        }
        \Else 
        {
            Choose the sequence from $\mathcal{S}_O$ using the following: $\boldsymbol{s}_F = \argmax\limits_{\boldsymbol{s} \in \mathcal{S}_O}{\mathlarger{\sum}\limits_{i \in F_1}^{}i} - {\mathlarger{\sum}\limits_{j \in F_2}j}$
        }
    }
\Else
{
    $\boldsymbol{s}_F \leftarrow \boldsymbol{s}^0$
}
\caption{\small{Algorithm for determining the sequence of cars in the SZ.}}\label{alg:cap}
\end{algorithm}

\subsubsection{\textbf{Lower level control: Decentralized motion control}}

Based on the sequence selected by the coordinator (i.e., the specific $(i^-,i^+)$ assigned to each CAV $i$), a decentralized low-level controller onboard each CAV $i$ adopts a control policy to follow the sequence. In particular, we use an MPC formulation to minimize Objectives 1,2 over $[t_i^0,t_i^{AZ}$, where $t_i^{AZ} < t_i^m$ is the time when $i$ enters the AZ and is treated differently (see Section \ref{sec:AZ}),
as well as satisfy Constraints 1-4. However, instead of formulating the MPC problem with the original constraints, we replace them with new ones based on CBFs.
There are two reasons for this approach: 1) Exploiting the forward invariance property of CBFs \cite{xiao2023safe}, the CBF-based control ensures that the original constraints are satisfied overall $t \in [t_i^0,t_i^{AZ}]$. 
2) Exploiting the fact that CBF constraints are \emph{linear} in the control, the computational cost of the MPC problem subject to CBFs is significantly lower and enables its online use.

We begin by deriving the CBF constraints ensuring that the constraints (\ref{VehicleConstraints1}), \eqref{rear-end} and \eqref{SafeMerging_tim} and (\ref{merging_2_tim}) are satisfied, subject to the vehicle dynamics in (\ref{VehicleDynamics}).
We define $f(\boldsymbol{x}_i(t))=[v_i(t),0]^T$ and $g(\boldsymbol{x}_i(t))=[0,1]^T$. 
Each of these constraints can be easily written in the form $b_q(\boldsymbol{x}(t)) \geq 0$, $q \in \lbrace 1,\ldots,n \rbrace$ where $n$ is the number of constraints and $\boldsymbol{x}(t)=[\boldsymbol{x}_1(t),\ldots,\boldsymbol{x}_{N(t)}(t)]$. The CBF method \cite{xiao2023safe}
maps a constraint $b_q(\boldsymbol{x}(t)) \geq 0$ onto a new constraint which is \emph{linear} in the control input and takes the general form
\begin{equation} \label{cbf_condition}
L_fb_q(\boldsymbol{x}(t))+L_gb_q(\boldsymbol{x}(t))u_i(t)+\alpha_q( b_q(\boldsymbol{x}(t))) \geq 0.
\end{equation}
where $L_f$, $L_g$ are the usual Lie derivatives along $f$ and $g$ respectively, and
$\alpha_q(\cdot)$ is an extended class $\mathcal{K}$ function, usually taken to be a linear function.

Starting with (\ref{VehicleConstraints1}), let $b_1(\boldsymbol{x}_i) := v_{max} - v_i(t)$ and $b_2(\boldsymbol{x}_i) = v_i(t) - v_{min}$. The corresponding CBF constraints are:
\begin{equation} \label{CBF1}
-u_i(t)+\alpha_1\big(b_1(\boldsymbol{x}_i)\big) \geq 0 
\end{equation}
\begin{equation} \label{CBF2}
u_i(t)+\alpha_2\big(b_2(\boldsymbol{x}_i)\big) \geq 0,  
\end{equation}
Letting $b_3(\mathbf{x}_i):= z_{i,i_p}(t) - \varphi v_{i}(t) - \delta$, the CBF constraint corresponding to the rear-end safety constraint in 
\eqref{rear-end} is
\begin{equation}\label{CBF3}
v_{i_p}(t)-v_i(t)-\varphi u_i(t)+\alpha_3(b_3(\textbf{x}_i)) \geq 0,
\end{equation}
Letting $b_4(\textbf{x}_i):=z_{i,i^+}(t_{i}^{m}) - \Phi(x_i(t)) v_{i}(t)+\delta$, the CBF constraint for safe merging in (\ref{SafeMerging}) is  
\begin{align} \label{CBF4}
& v_{i^+}(t)-v_i(t)- \frac{\partial \Phi\left(x_{i(t)}\right)}{\partial x_{i}} {v}_{i}(t) -\Phi\left(x_{i}(t)\right) u_i(t)+\\ \nonumber
&\alpha_4(b_4(\textbf{x}_i)) \geq 0
\end{align} 

Finally, let $b_5(\boldsymbol{x}_i):=x_{i}(t_{i}^{m}) - x_{i^-}(t_{i}^{m}) - \Phi(x_{i^-}(t_{i}^m)) v_{i^-}(t_{i}^{m})-\delta $. 
In this case, the relative degree of $b_5(\boldsymbol{x}_i)$ is greater than one (recall that the relative degree of a differentiable function $b: \mathbb{R}^n \rightarrow \mathbb{R}$ with respect  to system \eqref{VehicleDynamics} is the number of times it needs to be differentiated along its dynamics until the control $\boldsymbol{u}$ explicitly appears in the corresponding derivative \cite{xiao2023safe}).
Therefore, a High Order CBF (HOCBF) $\psi_1(\boldsymbol{x}_i)$ \cite{xiao2023safe} is needed and derived as follows:
\begin{align}
    \psi_1(\boldsymbol{x}_i) = v_i(t) - v_{i^-}(t) - \frac{\partial \Phi (\boldsymbol{x}_i)}{\partial \boldsymbol{x}_i}v_{i^-}(t)-\Phi (\boldsymbol{x}_i) u_{i^-} + \alpha_5\left(\boldsymbol{x}_i\right) \nonumber
\end{align}
\begin{align} \label{CBF5}
& u_{i}(t)-u_{i^-}(t)- \frac{\partial^2 \Phi\left(x_{i}(t)\right)}{\partial x_{i}^2} {v}_{i^-}(t) - 2\frac{\partial \Phi\left(x_{i}(t)\right)}{\partial x_i} {u}_{i^-}(t) \\ \nonumber 
& - \Phi\left(x_{i}(t)\right) \dot{u}_{i^-}(t) - \frac{d\alpha_5\left(b_{5}(t)\right)}{dt} + \alpha_6(\psi_1(\boldsymbol{x}_i))\geq 0
\end{align} 
Observe that the terms $u_{i^-}$ and $\dot{u}_{i^-}$ appear in this CBF constraint.   

In addition, we use a Control Lyapunov Function (CLF) \cite{xiao2023safe} $V(\cdot)$ associated with driving the CAV speed to a desired value $v_{i}^{ref}(t)$ by setting $V(\boldsymbol{x}_i(t))=\big(v_i(t)-v_{i}^{ref}(t)\big)^2$. The corresponding CLF constraint is
\begin{equation}\label{CLF_constraint}
L_fV\big(\boldsymbol{x}_i(t)\big)+L_gV\big(\boldsymbol{x}_i(t)\big)\boldsymbol{u}_i(t)+ c_3 V\big(\boldsymbol{x}_i(t)\big)\leq e_i(t),
\end{equation}
where $e_i(t)$ is a controllable relaxation variable which makes this a soft constraint, i.e., we do not require that $v_{i}(t)$ attains the desired value but only tracks it as closely as possible.


\textbf{MPC-CBF control problem} : We discretize $[t_i^0,t_i^{AZ}]$ using time steps of equal length $T_d$. The dynamics in \eqref{VehicleDynamics} are also discretized and we set $\boldsymbol{x}_{i,k}=\boldsymbol{x}_i(t_{i,k})$ where $k=1,2,\ldots$ is the $k$th time step. The decision variables $u_{i,k}=u_i(t_{i,k})$ and $e_{i,k}=e_i(t_{i,k})$ are assumed to be constant over each time step. Thus, the MPC optimization problem with CBF constraints is
\begin{align} \label{eqn:MPC_CBF_CLF}
&\min_{\boldsymbol{x}_{i,k}(\cdot),\boldsymbol{u}_{i,k}(\cdot), \boldsymbol{e}_{i,k}(\cdot)}  \sum_{h=0}^{H-1} \big({u}_{i,k+h} - u_{i}^{ref}(t_{i,k+h})\big)^2 + \beta_1 e_{i,k+h}^2 \\ \nonumber 
    \textnormal{s.t.} &  \ \   \boldsymbol{x}_{i,k+h+1}=f(\boldsymbol{x}_{i,k+h},\boldsymbol{u}_{i,k+h}), ~h=0,...,H-1 \\
    &\eqref{CBF1}, \eqref{CBF2}. \eqref{CBF3},\eqref{CBF4}, \eqref{CBF5}, \eqref{CLF_constraint}, \eqref{VehicleConstraints2} \nonumber \\
    &\boldsymbol{x}_{i,k}=\boldsymbol{x}_i(t_{i,k}) \nonumber     
\end{align}
where $\boldsymbol{u}_{i,k} = [u_{i,k},...,u_{i,k+H-1}], \boldsymbol{e}_{i,k} = [e_{i,k},...,e_{i,k+H-1}]$,
$H$ is the number of prediction horizon steps, $\beta_1$ is a weight factor for the CLF constraint in \eqref{CLF_constraint}, and $\boldsymbol{x}_i(t_{i,k})$ is given. Note that, in case, $i^+$ (or, $i^-$) $\in \emptyset$, then constraints \eqref{CBF4} (or, \eqref{CBF5}) are not applicable to \eqref{eqn:MPC_CBF_CLF}. Notice $i^- \in \{\emptyset, CAV\}$; we set $u_{i^-,k+h+1}=u_{i^-}(t_{i,k}) \ \forall h$ which makes $\dot{u}_{i^-} = 0$ for \eqref{CBF5}.  

The resulting control from \eqref{eqn:MPC_CBF_CLF} aims at tracking a given reference control trajectory $u_{i}^{ref}(t_{i,k})$, $k=1,2,\ldots$ .This is selected as the solution of an optimal control problem, presented next, which jointly minimizes the remaining travel time $t_i^{AZ}-t_{i,k}$ of CAV $i$ and its energy, for a fixed final state and given initial state:


\textbf{Problem $P(v_i^f)$}:
\begin{align} \label{eqn:Fixed_terminal_state}
 &\min_{t_i^{AZ}, {u_i}(.)}  \int_{t_{i,k}}^{t_i^{AZ}} \frac{1}{2}{u_i^2}(\tau) d\tau \\ \nonumber
 \textnormal{s.t.}& \ \ \
    x_i(t_i^{AZ})  = L - L_{AZ}, ~v_i(t_i^{AZ}) = v_i^f; ~ \textnormal{given} \ \ \boldsymbol{x}_i(t_{i,k}) 
\end{align}
where $v_i^f$ is the desired terminal velocity assigned to CAV $i$ when reaching the AZ; this is assigned by the lower level controller depending on the conditions it observes at each time $t$ as described next. The initial condition $\boldsymbol{x}_i(t_{i,k})$ is the state of CAV $i$ as evaluated by the solution of (\ref{eqn:MPC_CBF_CLF}) at the last time step ($k=1$ corresponds to the CAV first entering the CZ).
The solution of (\ref{eqn:Fixed_terminal_state}) is easily obtained analytically through standard Hamiltonian analysis, similar to such problems solved in Chapter 9 of \cite{xiao2023safe}.


In view of (\ref{eqn:Fixed_terminal_state}) and (\ref{eqn:MPC_CBF_CLF}), the lower-level decentralized controller performs the following steps at time $t_{i,k}$.

\textbf{Step 1: CAV $i$ receives $i^-$ and $i^+$ and their state information from the coordinator.} 
The state values for relevant vehicles $i^-$ and $i^+$ (i.e. position and velocity) are sent by the coordinator to each CAV $i \in F^{SZ}$.

\textbf{Step 2: Selection of controller mode.}
A ``mode'' is selected by adjusting the form of the safe merging CBF constraints \eqref{CBF4} and \eqref{CBF5} in \eqref{eqn:MPC_CBF_CLF} in response to the most recent observed HDV behavior which affects the new safe sequence determined at $t$. There are three modes that the controller selects from that stems as a result of resequencing requiring the CAV to adjust its motion, mentioned below.\\ 
\textbf{1) Jump ahead: $i^+(t_{i,k}) < i^+(t_{i,k-1})$.} 
    As a result of a HDV accelerating/decelerating and changing its order in the most recent safe sequence, the new (still safe) sequence $\boldsymbol{s}(t_{i,k})$ places CAV $i$ ahead of the CAV it was previously behind. Therefore, the controller checks the constraint \eqref{SafeMerging_2} and switches to this mode upon detecting its violation. The controller sets $v^f_i = v_{max}$ so as to allow CAV $i$ to regain satisfaction of \eqref{SafeMerging_2} as soon as possible. 
    CAV $i$ solves $P(v_{max})$ to obtain $u^{ref}_{i}(\tau)$, $\tau \in [t_{i,k},t_i^{AZ}]$. Then, the constraint \eqref{CBF4} is introduced with $i^+(t_{i,k})$ in \eqref{eqn:MPC_CBF_CLF}. Finally, \eqref{CBF5} is removed from \eqref{eqn:MPC_CBF_CLF} to avoid infeasibility however it will be added once the sequence successfully takes place (i.e. the corresponding original constraint \eqref{SafeMerging_2} becomes positive).  For example in Fig. \ref{fig:merging}, if $\boldsymbol{s}^0(t_{i,k})$ becomes unsafe due to a change in the behavior of HDV $5$, then new sequence will be generated in which the jump ahead mode will be activated for CAV 6, $6^+(t_{i,k-1}) = 4 > 6^+(t_{i,k}) = 2 $. In case a CAV is unable to adjust to the resequencing resulting from the new $i^+(t_{i,k})$ to satisfy the constraint \eqref{CBF5} prior to reaching the AZ, then it can activate
    the \emph{Yield Mode} within the AZ (see Section \ref{sec:AZ}) to complete the merging safely. \\
\textbf{2) Fall behind: $i^+(t_{i,k}) > i^+(t_{i,k-1})$.} 
    In this case, the new sequence places the CAV so as to follow a vehicle than it was planning to merge behind. 
    This causes a violation of constraint \eqref{SafeMerging}. 
    The controller sets $v^f_i = v_{min}$ and solves $P(v_{min})$. Then, \eqref{CBF5} is modified with $i^-(t_{i,k})$ and \eqref{CBF4} is removed to avoid infeasibility in \eqref{eqn:MPC_CBF_CLF}. As an example, in Fig. \ref{fig:merging}, if the HDV $5$ driving behavior changes, this may cause an unsafe sequence. As a result, the upper-level controller adjusts through resequencing and generates a new safe sequence where CAV $4$ falls behind by assigning $4^+(t_{i,k-1}) = 3 < 4^+(t_{i,k})= 6$. \\
\textbf{3) Retain sequence.} 
This mode occurs either when resequencing does not affect CAV $i$ or after the CAV successfully jumps ahead (or jumps behind), i.e. satisfies \eqref{CBF4} (or \eqref{CBF5}). Thus, both of them are included in problem \eqref{eqn:MPC_CBF_CLF}. In this case, $u^{ref}_{i} = 0$ which simply implies that CAV $i$ maintains its current velocity. An example of resequencing in Fig. \ref{fig:merging} occurs when  $3^+(t_{i,k-1}) = 3^+(t_{i,k}) = 2$.

\begin{figure}[t] 
\centering
\includegraphics[scale=0.65]{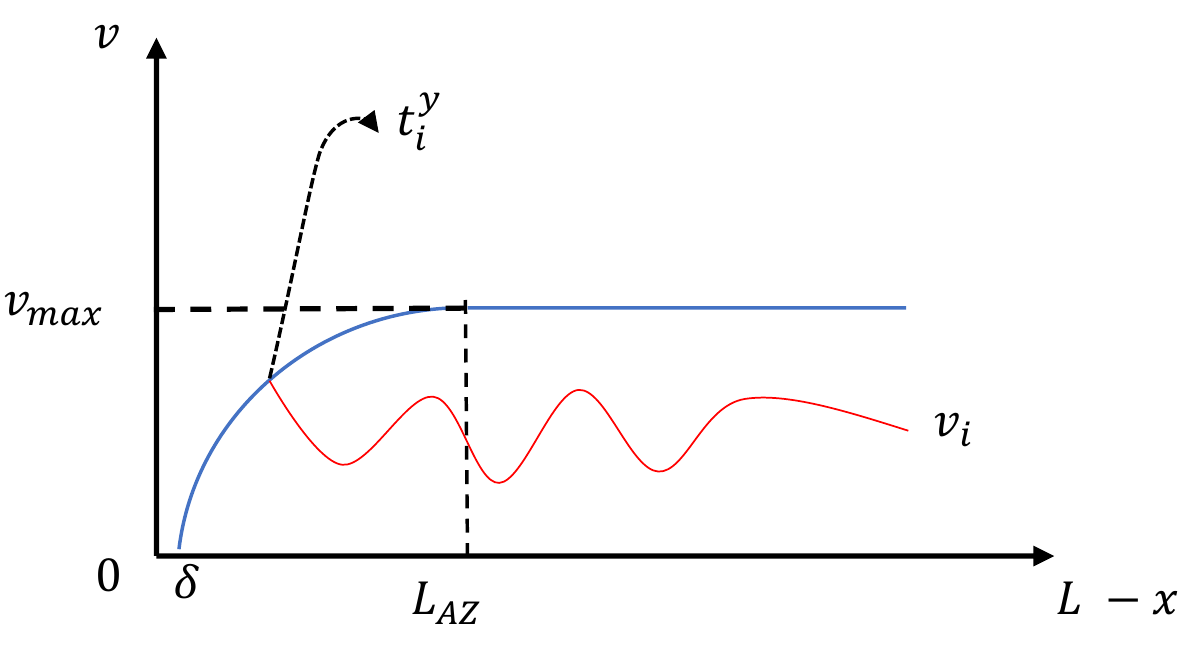} 
\caption{Maximum allowed velocity vs distance from $M$ to safely stop at $M$ with $u = u_{min}$}
\label{fig:velProfile}
\end{figure}


\textbf{Step 3: Obtain control input by solving the MPC-CBF control problem \eqref{eqn:MPC_CBF_CLF}.} 
After determining $u^{ref}_{i}$ in \textbf{Step 2}, $u_i(t_{i,k})$ can be obtained by solving \eqref{eqn:MPC_CBF_CLF} with proper constraints \eqref{CBF4} and \eqref{CBF5}. 

\section{Control in the Awareness zone} \label{sec:AZ}

The sequencing controller does not operate on vehicles located in the AZ since this is too short to make resequencing adjustments; moreover, the fact that vehicles can detect each other in the AZ facilitates interactions.
Although in many cases upon arrival of CAV $i$ at the AZ it can simply continue to follow the last safe sequence, the unpredictability of HDVs can also lead to potentially unsafe situations. As an example, suppose $i^- = \emptyset $ and a HDV chooses to be aggressive so as to become $i^-$ just upon arriving at the AZ and seeing the CAV. 
To account for such cases, we adopt a safe merging strategy for CAV $i$ which depends on its location in the AZ. 
As a last resort, we adopt a ``yield mode'' whereby the CAV simply yields to a conflicting HDV at the MP using an optimal trajectory that jointly minimizes its travel time and energy while ensuring that it allows the HDV to safely precede it.

Before proceeding, we introduce the notion of an \emph{aggresiveness parameter} for HDVs, used so that CAVs in the AZ can assess and predict the behavior of a HDV in terms of safely merging with them.


\textbf{HDV Aggressiveness parameter}: We define a parametric function that gives a measure of aggressiveness of a HDV at time $t_{i,k}$ based on its the state and input history (trajectory) since $t_{i}^{AZ}$ (time of arrival of CAV $i$ in AZ) of the form   
\begin{align} \label{aggressiveness_measure}
&a_{i^-} = f_{\theta}\big(\boldsymbol{x}_{i^-}(t_i^{AZ}), \dots, \boldsymbol{x}_{i^-}(t), \boldsymbol{u}_{i^-}(t_i^{AZ}), \dots, \boldsymbol{u}_{i^-}(t)\big)  
\end{align}
where $a_{i^-} \in [0,1]$ with higher values implying that the HDV is more aggressive. We assume that such a measure can be obtained through the application of a standard learning-based technique that uses the data in (\ref{aggressiveness_measure}) as its input.

{\textbf{Decentralized control}}
We consider a CAV $i \in F_r(t)$ with $i^- \in F_H(t)$ upon arriving at the AZ. Figure \ref{fig:velProfile} shows the relationship between the maximum velocity and the distance required to come to a full stop at the MP $M$ under maximum deceleration $u_{min}$ so as to highlight the existence of a critical time before yield mode $t^y_i$ when it is feasible for CAV $i$ to achieve this full stop. The value of $t^y_i$ depends on the aggressiveness $a_{i^-}$. Thus, the controller for CAV $i$ selects one of two modes:
$(i)$ If the aggressiveness parameter is less than a preset threshold $\gamma$, then it solves problem \eqref{eqn:MPC_CBF_CLF} to safely merge ahead of the HDV, or
$(ii)$ Otherwise, it comes to a complete stop at $M$ yielding to the HDV and using the \emph{Yield Mode} described next.

\textbf{Yield Mode:} In this case, CAV $i$ yields to make room for the HDV by solving the following fixed terminal state optimal control problem:
\begin{align} \label{eqn:MPC_CBF_CLF_Stop}
 &\min_{t_i^m, u_i(.)}  \int_{t_i^y}^{t_i^m}  \frac{1}{2} {u_i^2}(\tau) d\tau \\ \nonumber
    \textnormal{s.t.} \ \ \
   & x_i(t_i^{m})  = L - \delta, ~v_i(t_i^y) = 0; ~~ \textnormal{given } \boldsymbol{x}_i(t_i^y)
\end{align}
where $\boldsymbol{x}_i(t_i^y)$ is the known initial state of CAV $i$. The solution of this problem provides $u_{i}^{ref}$ for problem \eqref{eqn:MPC_CBF_CLF} subject to \eqref{CBF1},\eqref{CBF2},\eqref{CBF3},\eqref{CBF4},\eqref{CBF5}, and \eqref{VehicleConstraints2}.

Finally, if CAV $i \in F_r(t)$ and $i^- \in F_{\bar{r}}(t) \cap F_C(t)$, then the two CAVs simply cooperate to safely merge based on their respective distance to the merging point.

\section{Simulation Results}
All simulations 
are based on \textsc{PTV Vissim}. All algorithms are
implemented using \textsc{MATLAB} and \textsc{ode45} to integrate the CAV dynamics and we used CasADi to solve the minimization problem in \eqref{eqn:MPC_CBF_CLF}.
We have simulated the merging problem shown in Fig. \ref{fig:merging} where vehicles arrive according to Poisson arrival processes with an arrival rate that is fixed for the purpose of comparing the proposed safe sequencing options. The  initial speed $v_{i}(t_i^0)$ is also randomly generated with a uniform distribution over $[60 \textnormal{km/h}, 100\textnormal{km/h}]$ at the origins $O_1$ and $O_2$, respectively.
The remaining parameters used are: $L= 400\textnormal{m}$, $L_{SZ} = 300\textnormal{m}, L_{AZ} = 100\textnormal{m}, \varphi = 1.8\textnormal{s}, \delta = 3.78 \textnormal{m}, \beta_1= 1, u_{\max} = 4.905 \textnormal{m/s}^2, u_{\min} = -5.886\textnormal{m/s}^2, v_{\max} = 108 \textnormal{km/h}, v_{\min} = 0 \textnormal{km/h}$,$H = 15$.

The arrival rate of CAVs with respect to HDVs is chosen such that results with different penetration rates are generated.
In our simulations, we also included the computation of a more realistic energy consumption model \cite{kamal2012model} to supplement the simple surrogate $L_2$-norm ($u^2$) model in our analysis:
$f_{\textrm{v}}(t)=f_{\textrm{cruise}}(t)+f_{\textrm{accel}}(t)$ with
\begin{align*}
    f_{\textrm{cruise}}(t) &= \omega_0+\omega_1v_i(t)+\omega_2v^2_i(t)+\omega_3v^3_i(t),\\
    f_{\textrm{accel}}(t) &=\big(r_0+r_1v_i(t)+r_2v^2_i(t)\big)u_i(t).
\end{align*}
where we used typical values for parameters $\omega_1,\omega_2,\omega_3,r_0,r_1$ and $r_2$ as reported in \cite{kamal2012model}.

Simulation results of 100 cars with different CAV penetration rates are shown in Table I with respect to two different sequencing options: the proposed safe sequencing (SS) and SDF. The performance of the network in terms of average travel time, average energy consumption ($L_2$ norm of the control input) and average fuel consumption are compared. 
The significance of these results is twofold: 
$(i)$ Increasing the CAV rate under conventional sequencing policies, such as SDF, can potentially worsen the network performance until the penetration rate exceeds a certain threshold, 
$(ii)$ Even at low penetration rates, the proposed SS scheme can improve the performance of the entire network while favoring safety. Therefore, even at low penetration rates, the benefits of CAVs can be exploited. 

\begin{figure}
\centering
\includegraphics[scale=0.4]{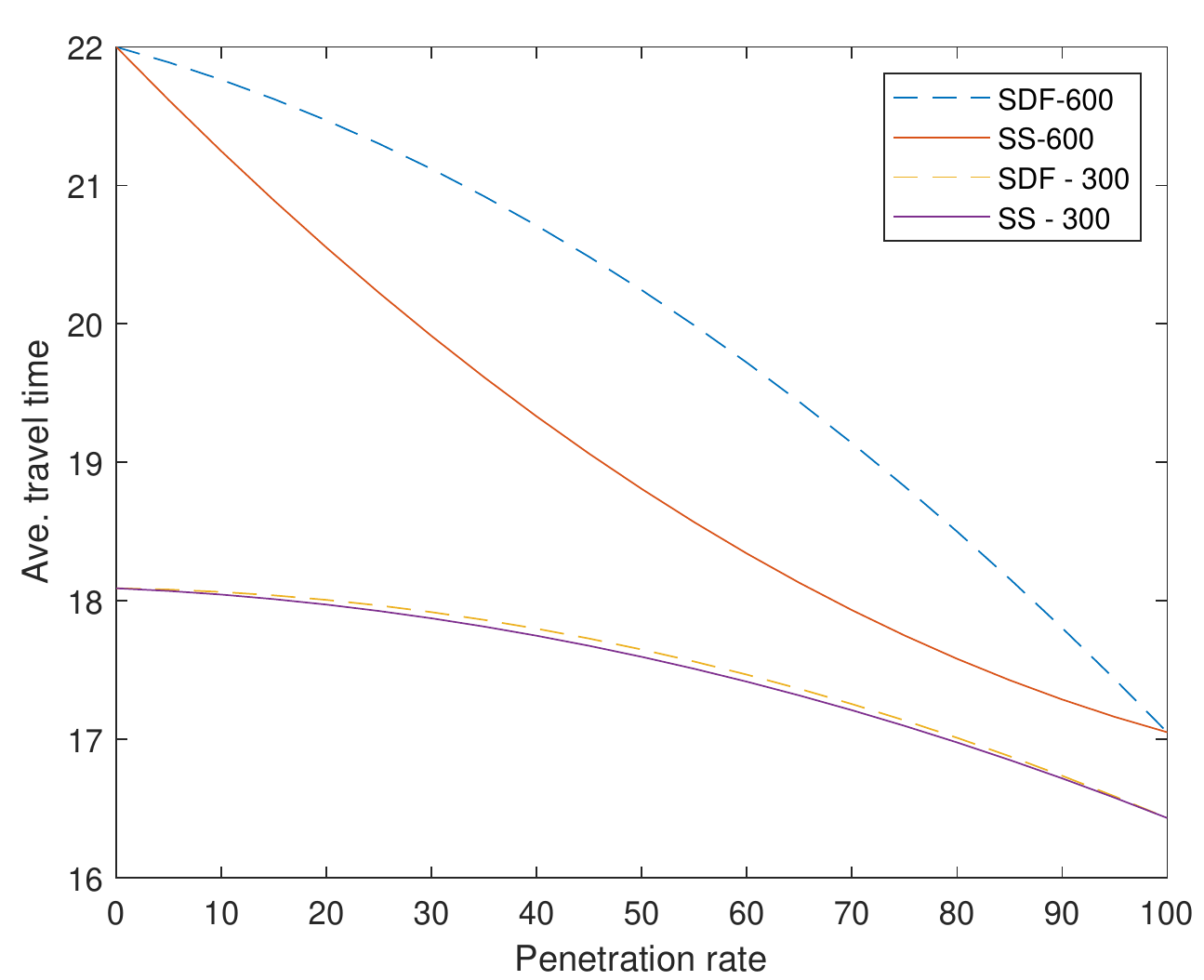} 
\caption{Average travel time of vehicles under SDF and SS}
\label{fig:SSvsSDF_time}
\end{figure}

\begin{figure}
\centering
\includegraphics[scale=0.4]{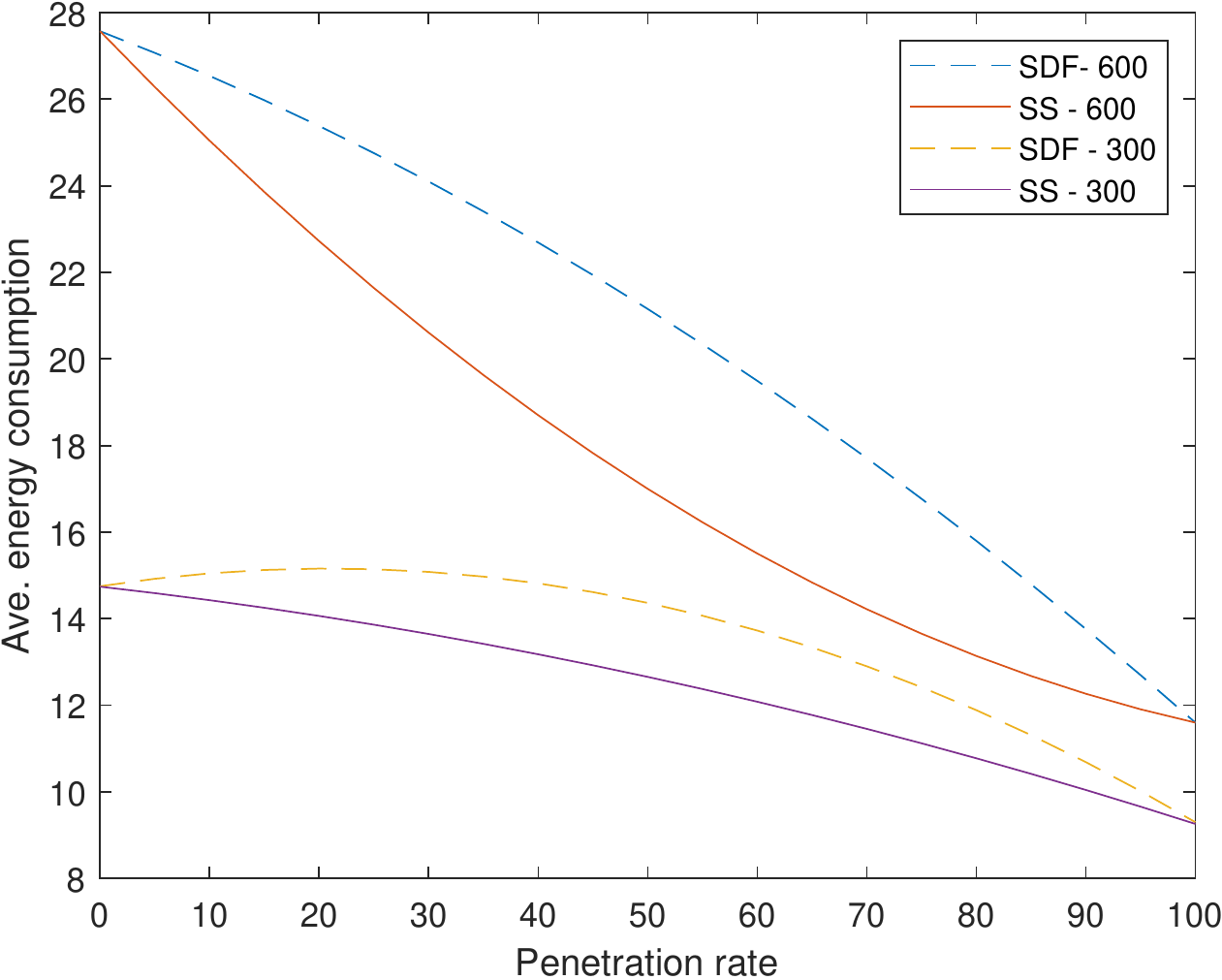} 
\caption{Average energy consumption of vehicles under SDF and SS }
\label{fig:SSvsSDF_energy}
\end{figure}

\begin{figure}
\centering
\includegraphics[scale=0.4]{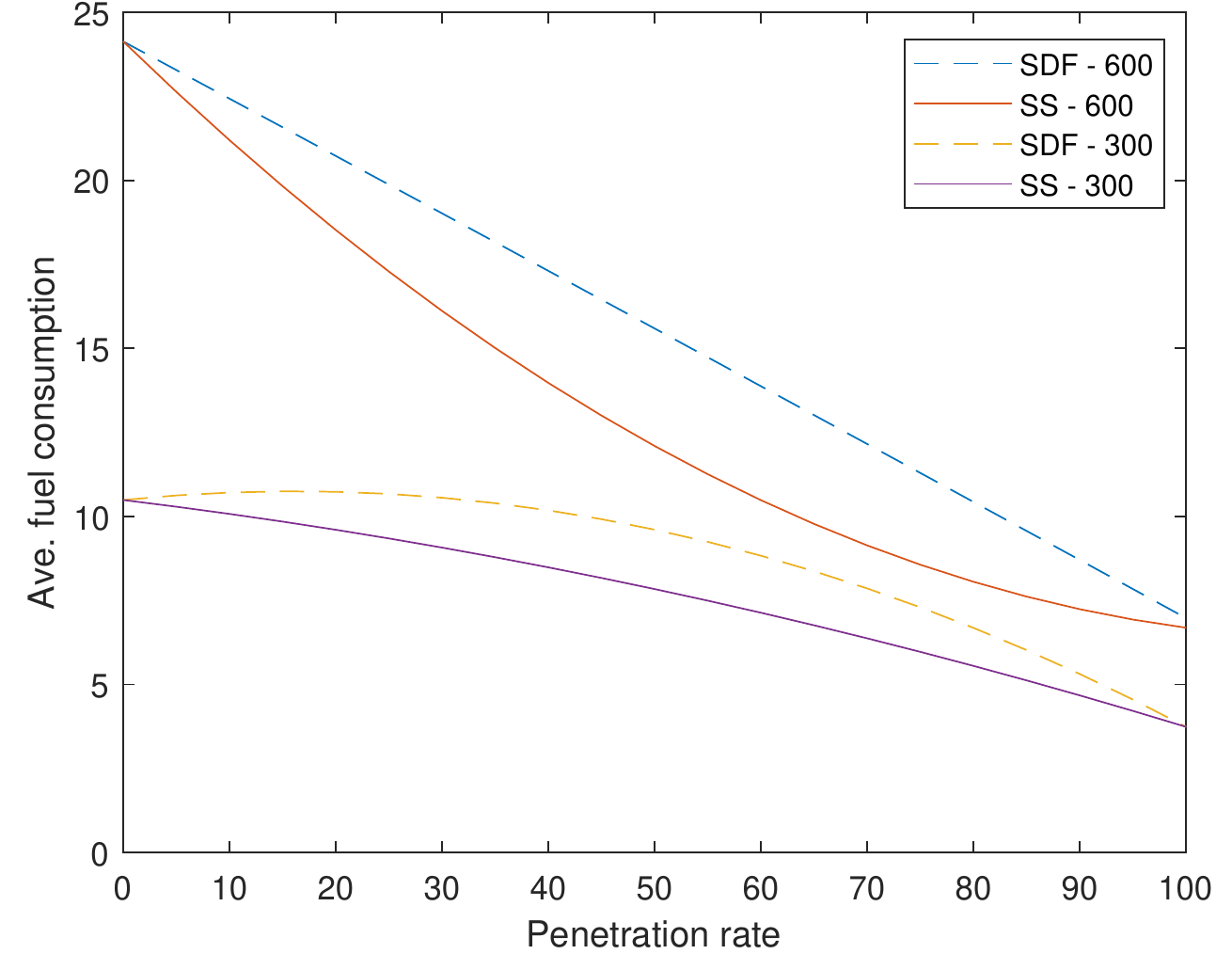} 
\caption{Average fuel consumption of vehicles under SDF and SS }
\label{fig:SSvsSDF_fuel}
\end{figure}

The performance metrics in Table I are illustrated in Figs. \ref{fig:SSvsSDF_time}, \ref{fig:SSvsSDF_energy} and \ref{fig:SSvsSDF_fuel}, where it can be clearly seen that SS uniformly outperforms conventional SDF sequencing while ensuring merging safety. In Fig. \ref{fig:figure 1} 
the results from multiple simulations are shown for a specific CAV penetration rate and fixed arrival rate for both SDF and SS. 

\begin{figure}
\centering
\includegraphics[scale=0.4]{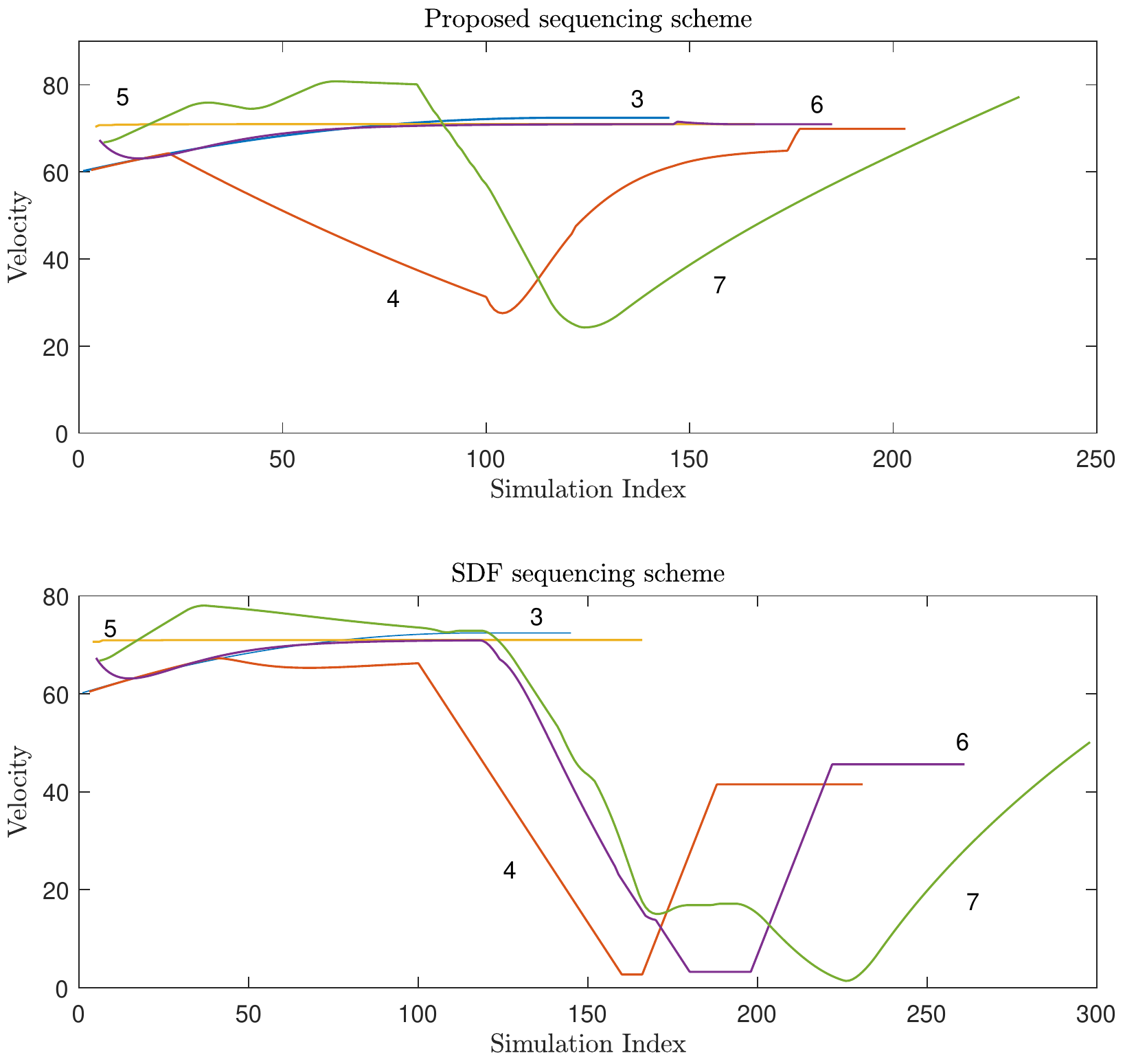} 
\caption{Velocity profile of vehicles present in the SZ under SDF and SS}
\label{fig:SSwithSDF}
\end{figure}

\begin{figure}
\centering
\includegraphics[scale=0.4]{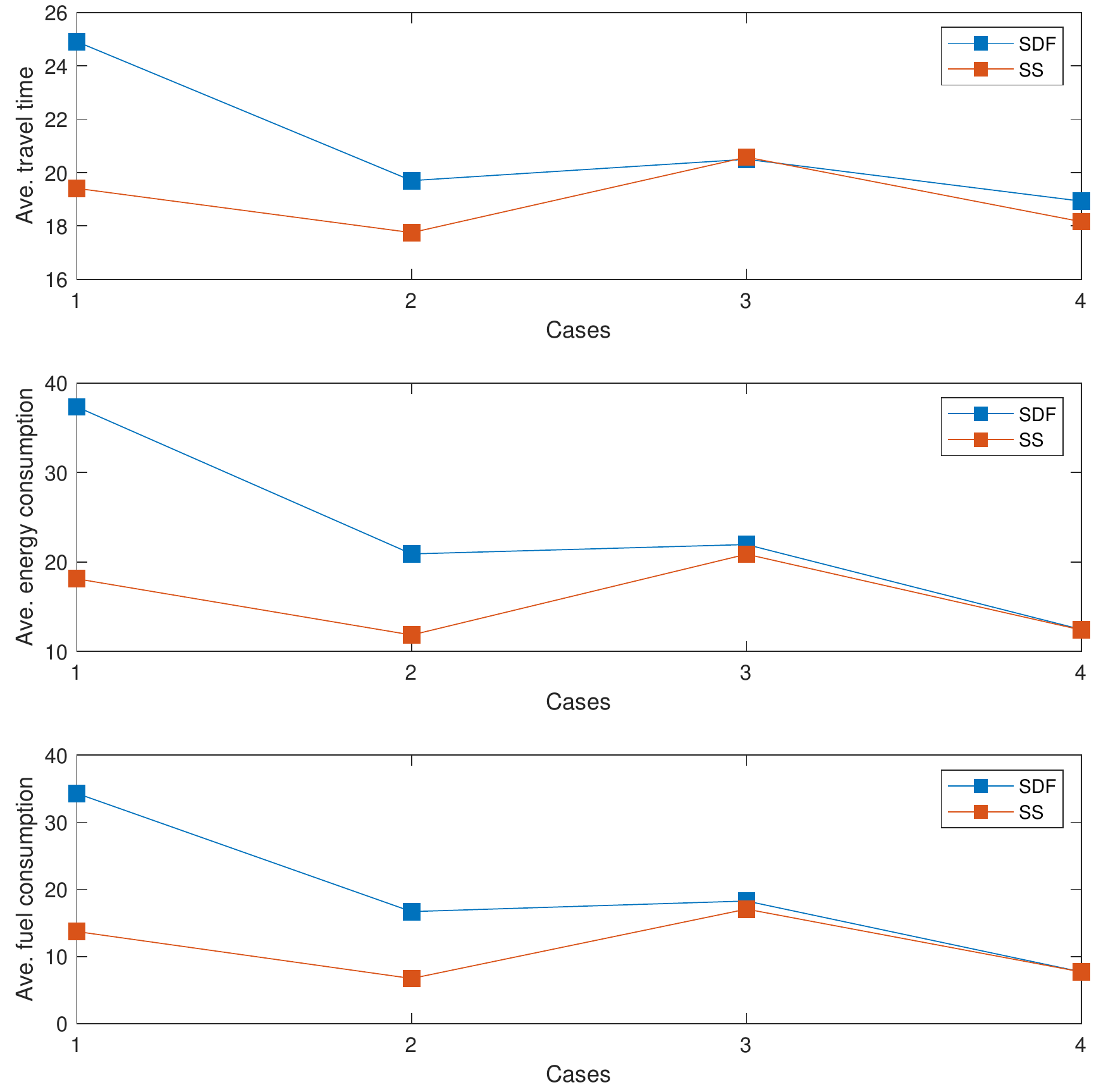} 
\caption{Comparison of SS and SDF schemes with fixed penetration rate and arrival rate of 600 $veh/hr$}
\label{fig:figure 1}
\end{figure}

One of the main reasons behind the uniform performance improvement we observe (despite the conservative nature of SS to guarantee safety) can be seen in Fig. \ref{fig:SSwithSDF} where 3 CAVs and 2 HDVs are present in the SZ as shown in Fig 1. Based on the current relative position of the vehicles, the SDF sequence is $\boldsymbol{s}^0=\{3,4,5,6,7\}$ which according to our threshold in \eqref{i-i+}, CAVs with indices 4 and 6 may need to be re-sequenced as car 5 and 7 are HDV. 
If we consider the case for which the safety constraints for both CAV 4 and CAV 6 is violated, i.e. $4^- = 5$ and $6^- = 7$, then based on the SS controller,  the new sequence will be $\boldsymbol{s}_F=\{3,5,6,4,7\}$. The speed profile of the cars with the new safe sequence is depicted in the lower plot in Fig. \ref{fig:SSvsSDF_time}. The main observation here is that the new safe sequence prevents excessive slowing down, especially in the main road. However, when we adopt SDF, after arriving in the AZ, CAV 4 decelerates rapidly coming to a complete halt to make room for HDV 5 since it does not yield to CAV 4, as illustrated in Fig. \ref{fig:SSwithSDF}. Similarly, since CAV 6 has to cooperate with CAV 4, it also decelerates rapidly to allow CAV 4 to merge ahead of it. Finally, as illustrated CAV 6 causes HDV 7 to decelerate rapidly yielding to CAV 6.

\begin{table}\scriptsize
        \centering
        \begin{tabular}{|c|c|c|c|}
            \cline{1-4}
             PR & Item & SS & SDF \\
        \hline  
        \multirow{3}{*}{{$0 \%$ }} & Ave. Travel time  & - & 17.71\\
        \cline{2-4}
        & Ave. $\frac{1}{2} u^2$ & - & 10.95\\
        \cline{2-4}
        & Ave. Fuel consumption & - &  5.86 \\        
        \hline        
        \multirow{3}{*}{{$ 20 \%$ }} & Ave. Travel time  &  \textcolor{blue}{17.5} & \textcolor{red}{18.5} \\
        \cline{2-4}
        & Ave. $\frac{1}{2} u^2$ & \textcolor{blue}{10.19} & \textcolor{red}{12.01} \\
        \cline{2-4}
        & Ave. Fuel consumption & \textcolor{blue}{5.01} & \textcolor{red}{5.89} \\
         \hline        
        \multirow{3}{*}{{$ 40 \%$ }} & Ave. Travel time  & \textcolor{blue}{17.16} & \textcolor{red}{19.2} \\
        \cline{2-4}
        & Ave. $\frac{1}{2} u^2$ & \textcolor{blue}{10.03} & \textcolor{red}{12.5}\\
        \cline{2-4}
        & Ave. Fuel consumption & \textcolor{blue}{4.66} & \textcolor{red}{6.79}\\
        \hline        
        \multirow{3}{*}{{$ 60 \%$ }} & Ave. Travel time  & \textcolor{blue}{16.90} & \textcolor{red}{18.33}\\
        \cline{2-4}
        & Ave. $\frac{1}{2} u^2$ & \textcolor{blue}{10.81} & \textcolor{red}{13.82}\\
        \cline{2-4}
        & Ave. Fuel consumption & \textcolor{blue}{5.7} & \textcolor{red}{8.66}  \\
        \hline        
        \multirow{3}{*}{{$ 80 \%$ }} & Ave. Travel time  & \textcolor{red}{16.59} & \textcolor{blue}{16.5}\\
        \cline{2-4}
        & Ave. $\frac{1}{2} u^2$ & \textcolor{blue}{9.44} & \textcolor{red}{11.30}\\
        \cline{2-4}
        & Ave. Fuel consumption & \textcolor{blue}{5.77} & \textcolor{red}{5.9} \\
        \hline        
        \multirow{3}{*}{{$ 100 \%$ }} & Ave. Travel time  & -  & 16\\
        \cline{2-4}
        & Ave. $\frac{1}{2} u^2$ & - &  9.07\\
        \cline{2-4}
        & Ave. Fuel consumption & - & 3.11 \\
        \hline
        \end{tabular}
        \caption{Vehicle metrics under Safe Sequencing (SS) and SDF with different CAV penetration rates.  }
        \label{Table self}
\end{table}

\textbf{Computational complexity}: Controller is comprised of two parts: i) the high level controller with the computational complexity of $O\bigl(N_1(N_2+N_1)\bigl)$ where $N_1$ and $N_2$ are the number of cars in road 1 and road 2, respectively. ii) the low level controller which solves an MPC control problem. The computational complexity of MPC is determined by $H$, which is the number of steps in the look ahead process. The control algorithm in this framework can be implemented in real time if the value for $H$ is chosen appropriately. The computational time to run this algorithm with Intel i5 8th Gen for $N_1 = 10$ and $N_2 = 5$ on average is $0.06 \textnormal{s} < T_d = 0.1\textnormal{s}$.
\section{Conclusion}
We have addressed the problem of merging traffic from two roadways consisting of both CAVs and HDVs so as to guarantee safety despite the unpredictability of HDV behavior. We developed a hierarchical controller which first determines a safe sequence and then a lower level decentralized motion controller for each CAV based on MPC and CBFs jointly minimizes travel time and energy subject to hard safety constraints dependent on the given safe sequence. Our simulation results show that this hierarchical controller outperforms the SDF sequencing policy in terms of time and energy over the full range of CAV penetration rates, while also providing safe merging guarantees.

\section{Appendix}
\textbf{Property 1:}
    The optimal safe sequence $\boldsymbol{s}_F$ satisfies all original rear-end safety constraints in \eqref{rear-end}.
\emph{Proof:} 
By construction, the SDF sequence $\boldsymbol{s}^0$ preserves the order of vehicles within their respective road, since the roads are single lanes and overtaking is not allowed. 
For any safe sequence $\boldsymbol{s}$ constructed from $\boldsymbol{s}^0$, it must must preserve the same order.
Therefore, to generate all admissible merging sequences, we consider $(F_1,<)$ and $(F_2,<)$ to be partially order sets (poset) \cite{ComplexityPoset2000}. As a result, $\boldsymbol{s}$ has to satisfy:
$(i)$ $|\boldsymbol{s}(t)| = |F^{SZ}_1(t) \cup F^{SZ}_2(t)|$ and
$(ii)$ $\forall p,l \in F^{SZ}_1(t), \ \ \  \boldsymbol{s}(p) < \boldsymbol{s}(l) \Leftrightarrow p < l$.

\textbf{Property 2:}
    There always exists a safe sequence $\boldsymbol{s}$.

\emph{Proof:} 
For any $\boldsymbol{s}^0$, given $i$ as the index of first CAV in the SZ (if exists) on road $r$, $\exists \boldsymbol{s}_S$ s.t. $\boldsymbol{s}_S \in \mathcal{S}_S$ as follows:
\begin{equation}
    {\boldsymbol{s}_S = \{\boldsymbol{s}^{0}i^+,i^-F^{SZ}_{\Bar{r}},iF^{SZ}_r\}}
\end{equation}
where $\boldsymbol{s}^{0}i^+$ are the cars physically ahead of $i^+$ in road $\bar{r}$ in sequence $\boldsymbol{s}^{0}$, $i^-F^{SZ}_{\Bar{r}}$ are the cars physically behind $i^-$ in road $\bar{r}$ in SZ and $iF^{SZ}_r$ are the cars physically following $i$ in road $r$ in the SZ. Given sequence $\boldsymbol{s}_S$, $\nexists i \in F^{SZ}_C$ s.t. $i^- \in F^{SZ}_H$. Therefore $s_S$ is always a safe sequence candidate in Algorithm 1 (if at least one CAV exists).
\end{document}